\documentstyle[12pt]{article}
\begin{document}

\baselineskip 20pt

\begin{flushright}
\begin{tabular}{l}
NEAP-54, July 1997
\end{tabular}
\end{flushright}

\vspace*{1.5cm}

\begin{center}

{\Large{\bf Flavor-Nonconservation and {\it CP}-Violation}}

{\Large{\bf with Singlet Quarks}} {\large{$ {}^1 $}}

\vspace{1cm}

{\large{Katsuji Yamamoto $ {}^2 $}}

\vspace{1cm}

{\it Department of Nuclear Engineering}

{\it Kyoto University, Kyoto 606-01, Japan}

\end{center}

\vspace{1cm}

\baselineskip 15pt

\begin{flushleft}
{\large{\bf abstract}}
\end{flushleft}

\bigskip
Some aspects are considered on the flavor-nonconservation
and $ CP $-violation arising from the quark mixings
with singlet quarks.  In certain models incorporating the singlet quarks,
the contributions of the quark couplings to the neutral Higgs fields
may become more significant than those of the neutral gauge interactions.
Then, they would provide distinct signatures for new physics
beyond the standard model
in various flavor-nonconserving and $ CP $-violating processes
such as the neutron electric dipole moment, $ D^0 $-$ {\bar D}^0 $,
$ K^0 $-$ {\bar K}^0 $, $ B^0 $-$ {\bar B}^0 $,
$ b \rightarrow s \gamma $, and so on.

\vspace*{2cm}

\begin{flushleft}
$ {}^1 $ Talk presented at ``Masses and Mixings of Quarks and Leptons",

{~~}March, 1997, Shizuoka, Japan

\bigskip
$ {}^2 $ e-mail address: yamamoto@nucleng.kyoto-u.ac.jp
\end{flushleft}

\newpage
\section{Introduction}

{~~~}

Some extensions of the standard model may be expected
in various points of view.  Among such possibilities,
electroweak models incorporating $ {\rm SU(2)}_W \times {\rm U(1)}_Y $
singlet quarks with electric charges $ 2/3 $ and $ - 1/3 $
have been investigated extensively in the literature
\cite{Br1,Ag1,B-M,q-Qmix,Br2,L-S,Br3,Rb,KY,Ag2,K-Y}.
In the presence of singlet quarks, various interesting issues
are presented phenomenologically.  In particular, some novel features
arise from the mixings between the ordinary quarks ($ q $)
and the singlet quarks ($ Q $):
The CKM unitarity in the ordinary quark sector is violated,
and the flavor changing neutral currents (FCNC's) are present
at the tree-level in both the gauge and Higgs interactions of the quarks.
The $ q $-$ Q $ mixings even involve new $ CP $-violating sources.
In this talk, we make further considerations on the flavor-nonconservation
and $ CP $-violation arising from the quark mixings with singlet quarks.
We present some relevant formulations for the quark mixings
and the FCNC's, which are useful for making more precise estimates
on the $ q $-$ Q $ mixing effects in the flavor-nonconserving
and $ CP $-violating processes
such as the neutron electric dipole moment (NEDM), $ D^0 $-$ {\bar D}^0 $,
$ K^0 $-$ {\bar K}^0 $, $ B^0 $-$ {\bar B}^0 $,
$ b \rightarrow s \gamma $, and so on.
Then, these formulations are applied, for instance,
for estimating the neutral Higgs contributions
to the NEDM and the $ D^0 $-$ {\bar D}^0 $ mixing
in the presence of significant mixing
between the top quark and singlet $ U $ quarks
\cite{K-Y}.
These estimates may be relevant for investigating the electroweak baryogenesis
with $ CP $-violating $ t $-$ U $ mixing.
\cite{baryogenesis,tUbaryogenesis}.
Some variants of the electroweak model incorporating singlet quarks
will also be discussed later, in particular,
by considering possible parametrizations of the quark mass matrices,
relative significance between the gauge and Higgs FCNC's
and the role of the singlet Higgs field.

\section{Quark masses and mixings with singlet quarks}

{~~~}

We here describe explicitly a specific version of electroweak model
incorporating singlet quarks, where the quark Yukawa couplings
are given by
\begin{eqnarray}
{\cal L}_{\rm Yukawa}
&=& -~u^c_0 \lambda_u q_0 H - u^c_0 ( f_U S + f_U^\prime S^\dagger ) U_0
- U^c_0 ( \lambda_U S + \lambda_U^\prime S^\dagger ) U_0
\nonumber \\
&{~}& -~d^c_0 \lambda_d V_0^\dagger q_0 {\tilde H}
- d^c_0 ( f_D S + f_D^\prime S^\dagger ) D_0
- D^c_0 ( \lambda_D S + \lambda_D^\prime S^\dagger ) D_0 ~+~{\rm h.c.}
\label{eqn:LYukawa}
\end{eqnarray}
with the two-component Weyl fields (the generation indices and the factors
representing the Lorentz covariance are omitted for simplicity).
Here $ q_0 = ( u_0 , V_0 d_0 ) $ represents the quark doublets
with a unitary matrix $ V_0 $,
and $ H = ( H^0 , H^+ ) $ is the electroweak Higgs doublet
with $ {\tilde H} = ( H^- , - H^{0 \dagger} ) $.
Suitable redefinitions among the $ q^c_0 $ and $ Q^c_0 $ fields
with the same quantum numbers has been made
to eliminate the $ U^c_0 q_0 H $ and $ D^c_0 q_0 {\tilde H} $ couplings
without loss of generality.
Then, the Yukawa coupling matrices $ \lambda_u $ and $ \lambda_d $
have been made diagonal by using unitary transformations
among the ordinary quark fields.
In this basis, by turning off the $ q $-$ Q $ mixings
with $ f_Q , f_Q^\prime \rightarrow 0 $,
$ u_0 $ and $ d_0 $ are reduced to the mass eigenstates,
and $ V_0 $ is identified with the CKM matrix.
The actual CKM matrix is slightly modified due to the $ q $-$ Q $ mixings,
as shown explicitly later.
A complex singlet Higgs field $ S $ is introduced
in the present model given by eq. (\ref{eqn:LYukawa})
to provide the singlet quark mass terms and the $ q $-$ Q $ mixing terms.
Some variants of the model will be considered later by noting
whether the singlet Higgs field is introduced or not.

The Higgs fields develop vev's,
\begin{equation}
\langle H^0 \rangle = \frac{v}{\sqrt 2} ~,~~
\langle S \rangle = {\rm e}^{i \phi_S} \frac{v_S}{\sqrt 2}~,
\label{eqn:vev}
\end{equation}
where $ \langle S \rangle $ may acquire a nonvanishing phase $ \phi_S $
due to either spontaneous or explicit $ CP $ violation
originating in the Higgs sector.
The quark mass matrices are produced with these vev's as
\begin{equation}
{\cal M}_{\cal U} = \left( \begin{array}{cc}
M_u & \Delta_{u{\mbox{-}}U} \\ 0 & M_U \end{array} \right) ~,~~
{\cal M}_{\cal D} = \left( \begin{array}{cc}
M_d & \Delta_{d{\mbox{-}}D} \\ 0 & M_D \end{array} \right) ~.
\label{eqn:MQmatrix}
\end{equation}
These quark mass matrices are diagonalized by unitary transformations
$ {\cal V}_{{\cal Q}_{\rm L}} $ and $ {\cal V}_{{\cal Q}_{\rm R}} $
($ {\cal Q} = {\cal U}, {\cal D} $) as
\begin{eqnarray}
{\cal V}_{{\cal U}_{\rm R}}^\dagger {\cal M}_{\cal U}
{\cal V}_{{\cal U}_{\rm L}}
& = & {\rm diag.} ( m_u , m_c , m_t , m_{U_1}, \ldots ) ~,
\nonumber \\
{\cal V}_{{\cal D}_{\rm R}}^\dagger {\cal M}_{\cal D}
{\cal V}_{{\cal D}_{\rm L}}
& = & {\rm diag.} ( m_d , m_s , m_b , m_{D_1}, \ldots )
\label{eqn:MQdia}
\end{eqnarray}
with quark mass eigenvalues given by
\begin{equation}
m_{q_i} = \frac{\lambda_{q_i} v}{\sqrt 2}
\left[ 1 + {\cal O} ( \epsilon_{q{\mbox{-}}Q}^2 ) \right] ~.
\label{eqn:qmass}
\end{equation}
Here the parameters $ \epsilon_{q{\mbox{-}}Q} \sim | f_Q | + | f_Q^\prime | $
represent the mean magnitudes of the $ q $-$ Q $ mixings.
The generalized CKM matrix is given by
\begin{equation}
{\cal V} = {\cal V}_{{\cal U}_{\rm L}}^\dagger
\left( \begin{array}{cc}
V_0 & 0 \\ 0 & 0 \end{array} \right) {\cal V}_{{\cal D}_{\rm L}}
= \left( \begin{array}{cc}
V & {*} \\ {*} & {*} \end{array} \right) ~.
\label{eqn:VCKM}
\end{equation}
Here the CKM matrix $ V $ is no longer unitary due to the $ q $-$ Q $ mixings.
It is found by determinig $ {\cal V}_{{\cal Q}_{\rm L,R}} $ perturbatively
that the CKM unitarity violation arises as
\begin{equation}
( V^\dagger V - {\bf 1} )_{ij} ~,~ ( V V^\dagger - {\bf 1} )_{ij}
\sim  ( m_{u_i} m_{u_j} / m_U^2 ) \epsilon_{u{\mbox{-}}U}^2
+ ( m_{d_i} m_{d_j} / m_D^2 ) \epsilon_{d{\mbox{-}}D}^2 ~,
\label{eqn:unitarity-violation}
\end{equation}
being related to the FCNC's coupled to the $ Z $ boson
\cite{Br1,Ag1,q-Qmix,Br2,KY,Ag2,K-Y}.
This relation ensures that the CKM unitarity violation
is sufficiently below the experimental bounds
\cite{ParticleData}
for reasonable ranges of the model parameters.

\section{FCNC's}

{~~~}

The FCNC's, which may include $ CP $ violating sources,
arise in both the gauge interactions and the neutral Higgs couplings.
We here consider these FCNC's, respectively.

\begin{flushleft}
{\large{\bf FCNC's in the gauge interactions}}
\end{flushleft}

It is straightforward to write down the quark gauge interactions
coupled to the $ Z $ bosons in terms of the quark mass eigenstates:
\begin{equation}
{\cal L}_{\rm NC}(Z)
= g_Z Z_\mu \left[
\sum_{{\cal Q}={\cal U},{\cal D}}
{\cal Q}^\dagger \sigma^\mu {\cal V}_Z ({\cal Q}) {\cal Q} ~
+ \sum_{{\cal Q}^c = {\cal U}^c , {\cal D}^c}
{\cal Q}^{c \dagger} \sigma^\mu
{\cal V}_Z ({\cal Q}^c) {\cal Q}^c \right] ~,
\end{equation}
where
$ g_Z = g/ \cos \theta_W $, $ g = e/ \sin \theta_W $,
and the coupling matrices are given by
\begin{eqnarray}
{\cal V}_Z ({\cal Q}) &=& {\cal V}_{{\cal Q}_{\rm L}}^\dagger
\left( \begin{array}{cc} I_Z (q_0) {\bf 1} & 0 \\
0 & I_Z (Q_0) {\bf 1}
\end{array} \right) {\cal V}_{{\cal Q}_{\rm L}}
= \left( \begin{array}{cc} V_Z (q) & {*} \\ {*} & {*} \end{array} \right) ~,
\\
{\cal V}_Z ({\cal Q}^c) &=& I_Z ({\cal Q}^c_0)
\left( \begin{array}{cc} {\bf 1} & 0 \\ 0 & {\bf 1}
\end{array} \right)
\label{eqn:VZ}
\end{eqnarray}
with $I_Z ({\cal F}) = I_3 ({\cal F})
- \sin^2 \theta_W Q_{\rm EM} ({\cal F}) $
[ $ {\cal F} = {\cal Q}_0 , {\cal Q}^c_0 $ ].
The FCNC's do not appear for the right-handed quarks
in the gauge interactions, since they have the same
$ {\rm SU(2)}_W \times {\rm U(1)}_Y $ quantum numbers
with $ I_Z({\cal Q}^c_0) = I_Z(q^c_0) = I_Z (Q^c_0) $.
The coupling matrices $ V_Z (q) $ for the left-handed quarks
are actually determined in respective models
depending on how the singlet quark mass terms and the $ q $-$ Q $ mixing
terms are provided, and how the Yukawa couplings and the quark mass
matrices are parametrized in certain bases.
In the present model given by eq. (\ref{eqn:LYukawa}),
the $ U^c_0 q_0 H $ and $ D^c_0 q_0 {\tilde H} $ couplings
have been rotated out by suitable redefinition among the right-handed
quarks.  Then, the quark mass matrices have the specific form
(\ref{eqn:MQmatrix}), respecting the relation
$ m_{q_i} \sim \lambda_{q_i} v $ as given in eq. (\ref{eqn:qmass}).
By making perturbative calculations with these quark mass matrices
the $ q $-$ Q $ mixing effects on the neutral gauge couplings
of the ordinary quarks are found as
\begin{equation}
V_Z (q)_{ij} - V_Z (q_0)_{ij}
\sim ( m_{q_i} m_{q_j} / m_Q^2 ) \epsilon_{q{\mbox{-}}Q}^2 ~,
\label{eqn:dVZ}
\end{equation}
where the $ V_Z (q_0) $ represents the usual neutral currents
in the absence of $ q $-$ Q $ mixings.
This modification on the neutral currents is related to
the unitarity violation given in eq. (\ref{eqn:unitarity-violation})
\cite{Br1,Ag1,q-Qmix,Br2,KY,Ag2,K-Y}.

\begin{flushleft}
{\large{\bf FCNC's in the neutral Higgs couplings}}
\end{flushleft}

The quark couplings to the neutral Higgs fields are extracted
from (\ref{eqn:LYukawa}) as
\begin{equation}
{\cal L}_{\rm NC}({\rm Higgs})
= - \sum_{{\cal Q} ; a = 0,1,2}
{\cal Q}^c \Lambda^{\cal Q}_a {\cal Q} \phi_a ~+~ {\rm h.c.}~,
\label{eqn:LYneutral}
\end{equation}
where $ \phi_0 $, $ \phi_1 $, $ \phi_2 $ are the mass eigenstates
of the neutral Higgs fields.
Then, the coupling matrices in eq.(\ref{eqn:LYneutral}) are given by
\begin{equation}
\Lambda^{\cal Q}_a = \frac{1}{\sqrt 2} {\cal V}_{{\cal Q}_{\rm R}}^\dagger
( {\mbox{\sf O}}_{a0} \Lambda_q + {\mbox{\sf O}}_{a1} \Lambda_{Q}^+
+ i {\mbox{\sf O}}_{a2} \Lambda_{Q}^- ) {\cal V}_{{\cal Q}_{\rm L}}
\label{eqn:Lam-a}
\end{equation}
with
\begin{equation}
\Lambda_u = \left( \begin{array}{cc}
\lambda_u & 0 \\
0 & 0
\end{array} \right) ~,~~
\Lambda_d = \left( \begin{array}{cc}
- \lambda_d V_0^\dagger & 0 \\
0 & 0
\end{array} \right) ~,~~
\Lambda_{Q}^\pm = \left( \begin{array}{cc}
0 & f_Q \pm f^\prime_Q \\
0 & \lambda_Q \pm \lambda^\prime_Q
\end{array} \right) ~.
\label{eqn:Lam-qQ}
\end{equation}
Here an orthogonal matrix $ {\mbox{\sf O}} $ is introduced
to parametrize the mass eigenstates of the neutral Higgs fields.
It is seen by making perturbative calculations
that the ordinary quark couplings to the neutral Higgs fields,
in particular, have specific generation dependence as
\begin{equation}
( \Lambda^{\cal Q}_a )_{ij}
\sim ( m_{q_j} / m_Q ) \epsilon_{q{\mbox{-}}Q}^2 ~.
\label{eqn:Lam-a-ij}
\end{equation}

\bigskip
\begin{flushleft}
{\large{\bf FCNC's ($ Z $) versus FCNC's (Higgs)}}
\end{flushleft}

As seen in eqs. (\ref{eqn:dVZ}) and (\ref{eqn:Lam-a-ij}),
the FCNC's coupled to the neutral Higgs fields
are of the first order of the relevant ordinary quark masses,
while the $ q $-$ Q $ mixing effects on the $ Z $ boson
couplings appear at the second order.
This implies that the FCNC's (Higgs) are more significant
than FCNC's ($ Z $) in this sort of models with the quark mass matrices
of the form given in eq.(\ref{eqn:MQmatrix}),
where the quark mass hierarchy is respected naturally
under the relation (\ref{eqn:qmass}).
Then, the neutral Higgs contributions
in various flavor-nonconserving and $ CP $-violating processes
are expected to serve as signals
for the new physics beyond the standard model.

\section{Higgs contributions to NEDM and $ D^0 $-$ {\bar D}^0 $
with $ u $-$ U $ mixings}

{~~~}

The neutral Higgs contribution to the NEDM was considered earlier
\cite{Br2}, claiming that the singlet Higgs mass scale
$ | \langle S \rangle | = v_S /{\sqrt 2} $
should be in the TeV region or larger.
On the other hand, in view point of electroweak baryogenesis
\cite{baryogenesis,tUbaryogenesis},
the mass scale of the singlet Higgs field to provide the $ CP $-violating
$ q $-$ Q $ mixings is desired to be comparable to the electroweak scale.
In order to clarify this apparently controvertial situation,
detailed analyses have been made recently \cite{K-Y},
showing that the NEDM becomes comparable to the experimental bound
for the singlet Higgs mass scale
$ v_S \sim {\rm several} \times 100 {\rm GeV} $
even in the presence of significant $ t $-$ U $ mixing.
We here describe these analyses briefly,
where the $ D^0 $-$ {\bar D}^0 $ mixing is also considered.

The total one-loop contribution of the neutral Higgs fields
to the $ u $ quark EDM is calculated by a usual formula
\begin{equation}
d_u ( \phi ) = - \frac{2 e}{3 (4 \pi )^2} \sum_a
\sum_{{\cal U}_K = u_i , U} \left\{
{\rm Im} \left [  ( \Lambda_a )_{1 K} ( \Lambda_a )_{K 1}
\right] \frac{m_{{\cal U}_K}}{m_{\phi_a}^2}
I ( m_{{\cal U}_K}^2 / m_{\phi_a}^2 ) \right\} ~,
\label{eqn:du-phi}
\end{equation}
where $ I(X) $ is a certain function of $ X $.
The effective Hamiltonian for the $ D^0 $-$ {\bar D}^0 $ mixing,
on the other hand, is obtained from the quark couplings
to the neutral Higgs fields (\ref{eqn:Lam-a}) as
\begin{equation}
{\cal H}_{\phi}^{\Delta c = 2}
= \sum_a \frac{1}{m_{\phi_a}^2} \left[ {\bar c}
\left\{ ( \Gamma_a^S )_{21}
+ ( \Gamma_a^P )_{21} \gamma_5 \right\} u \right]^2 ~,
\label{eqn:H-phi}
\end{equation}
where
\begin{equation}
( \Gamma_a^S )_{21} = \frac{1}{2} \left[ ( \Lambda^{\cal U}_a )_{12}^*
+ ( \Lambda^{\cal U}_a )_{21} \right] ~,~~
( \Gamma_a^P )_{21} = \frac{1}{2} \left[ ( \Lambda^{\cal U}_a )_{12}^*
- ( \Lambda^{\cal U}_a )_{21} \right] ~.
\label{eqn:Gamma-SP}
\end{equation}
Systematic analyses have been done in \cite{K-Y}
for the neutral Higgs contributions (\ref{eqn:du-phi}) and (\ref{eqn:H-phi})
to the NEDM and the $ D^0 $-$ {\bar D}^0 $ mixing
due to the $ u $-$ U $ mixings.  There, the quark mass matrices
are diagonalized numerically to determine presicely the quark mixing matrices
and the quark couplings to the $ Z $ boson and neutral Higgs fields.
The relevant coupling parameters are taken in certain reasonable ranges
as
\[
\begin{array}{l}
\epsilon_{u{\mbox{-}}U} \sim 0.1 ~,~~
{\mbox{complex phases in $ {\cal L}_{\rm Yukawa} $
and $ \langle S \rangle $}} \sim 1 ~,
\\ ~~~ \\
m_U \sim {\rm several} \times 100 {\rm GeV} ~,
\\ ~~~ \\
m_{\phi_0} \sim 100 {\rm GeV} ~,~~
m_{\phi_1} , m_{\phi_2} \sim v_S
{\mbox{~$ > $ \hspace{-1.05em}{\raisebox{-0.75ex}{$ \sim $}}~}}
100 {\rm GeV} ~.
\end{array}
\]
The results are given as
\[
| d_u ( \phi ) | \sim ( 10^{-25} - 10^{-27} ) e{\rm cm} ~,~~
\Delta m_D ( \phi ) \sim ( 10^{-13} - 10^{-15} ) {\rm GeV} ~,
\]
which are comparable to the present experimental bounds
\cite{ParticleData}.

As for the case of $ d $-$ D $ mixings, the neutral Higgs contributions
to the $ K^0 $-$ {\bar K}^0 $ mixing and the NEDM should be
investigated as well.  It is actually found that the $ CP $ violation parameter
$ \epsilon $ for the $ K^0 $-$ {\bar K}^0 $ mixing, in particular,
can be as large as $ 10^{-3} $
for $ \epsilon_{d{\mbox{-}}D} \sim 0.1 $ and $ v_S \sim v $.

\section{Some variants of the model with singlet quarks}

{~~~}

We finally consider some possible variants of the model
incorporating the singlet quarks.

\begin{flushleft}
{\large{\bf alternative form of the quark mass matrices}}
\end{flushleft}

We have rotated out the $ u^c_0 q_0 H $ and $ d^c_0 q_0 {\tilde H} $
couplings in eq. (\ref{eqn:LYukawa}) providing the specific form
of the quark mass matrices (\ref{eqn:MQmatrix}).
This respects naturally the mass hierarchy of the ordinary quarks
with the relation (\ref{eqn:qmass}).
It is instead possible to elimiate the $ u^c_0 U_0 S $
and $ d^c_0 D_0 S $ couplings with $ f_U , f_D \rightarrow 0 $,
while the $ f_U^\prime $ and $ f_D^\prime $ couplings
may still have generic forms with the diagonal $ \lambda_u $
and $ \lambda_d $ couplings.
Then, alternative form of the quark mass matrices are obtained.
Even in this case, if the $ f_U^\prime $ and $ f_D^\prime $ couplings
are small enough (not necessarily for the top quark), then
the masses of the ordinary quarks are not changed significantly,
maintaining the natural relation $ m_{q_i} \sim \lambda_{q_i} v $.
It is interesting in this case, as confirmed by numerical calculations,
that the FCNC's ($ Z $) can be larger than the FCNC's (Higgs).
Then, for instance, a significant
contribution of the FCNC's ($ Z $) to the $ D^0 $-$ {\bar D}^0 $ mixing
may be obtained, as investigated in
\cite{Br3}.

\begin{flushleft}
{\large{\bf real singlet Higgs field}}
\end{flushleft}

The complex Higgs field $ S $ may be replaced by a real field
with $ f_Q^\prime = \lambda_Q^\prime = 0 $.
Even in this case, similar contributions are expected from the FCNC's (Higgs)
for the flavor-nonconserving and $ CP $-violating processes.
It should here be mentioned that with only one real singlet Higgs field
the $ t $-$ U $ mixing is ineffective for the electroweak baryogenesis.
This is because the complex phases in the $ t $-$ U $ couplings
to the real singlet Higgs field are eliminated away by rephasing
the $ U_0 $ and $ U_0^c $ fields.
The alternative form of the quark mass matrices,
\[
{\cal M}_{\cal Q} = \left( \begin{array}{cc}
M_q & 0 \\ \Delta^\prime_{q{\mbox{-}}Q} & M_Q \end{array} \right) ~,
\]
are also possible, as mentioned above, by redefining
the right-handed quark fields.

\begin{flushleft}
{\large{\bf no singlet Higgs field}}
\end{flushleft}

The singlet Higg field $ S $ may be absent with explicit mass terms
$ \Delta_{u{\mbox{-}}U} $ and $ M_U $ in eq. (\ref{eqn:MQmatrix}).
Even in this case, the significant FCNC's with $ CP $-violating phases
may still be present in the quark couplings to the $ Z $ boson
and the standard neutral Higgs field.
The one-loop neutral Higgs contribution to the NEDM is, however, vanishing,
just as the one-loop $ Z $ boson contribution.
This is because the standard neutral Higgs field
and the Nambu-Goldstone mode couple to the quarks
in the same way.

\section{Summary}

{~~~}

Some aspects have been considered on the flavor-nonconservation
and $ CP $-violation arising from the quark mixings with singlet quarks.
In certain models incorporating the singlet quarks,
the contributions of the quark couplings to the neutral Higgs fields
may become more significant than those of the neutral gauge interactions.
Then, they would provide distinct signatures for new physics
beyond the standard model
in various flavor-nonconserving and $ CP $-violating processes
such as the NEDM, $ D^0 $-$ {\bar D}^0 $,
$ K^0 $-$ {\bar K}^0 $, $ B^0 $-$ {\bar B}^0 $,
$ b \rightarrow s \gamma $, and so on.
It is, for instance, found that the neutral Higgs contributions
to the NEDM and the neutral $ D $ meson mass difference
can be comparable to the present experimental bounds
for the case where the singlet Higgs mass scale is of order
of the electroweak scale and a significant $ CP $-violating
$ t $-$ U $ mixing is present.
This situation may be desired for the electroweak baryogenesis
with $ t $-$ U $ mixing.

\bigskip
\begin{flushleft}
{\large{\bf acknowledgement}}
\end{flushleft}

I would like to thank I. Kakebe for his collaboration
in a part of this work.

\vspace*{1.0cm}

\end{document}